\documentclass[11pt]{amsart} 
\usepackage{curves}
\usepackage{euler}
\usepackage{euscript}
\usepackage{multicol} 
\usepackage{amssymb}
\usepackage{amsmath} 
\usepackage{times} 

\newcommand{\tr}{\mbox{\upshape tr\ }}
\newtheorem{theorem}{Theorem}[section] 
\newtheorem{lemma}[theorem]{Lemma}
\newtheorem{definition}[theorem]{Definition}
\newtheorem{corollary}[theorem]{Corollary}

\numberwithin{equation}{section}

\begin{document}
\title[Lieb-Thirring inequalities]{Lieb-Thirring 
Inequalities for Higher Order Differential Operators}
\author[C. F\"{o}rster, J. \"{O}stensson]{Clemens F\"{o}rster and J\"{o}rgen \"{O}stensson}
\begin{abstract}
We derive Lieb-Thirring inequalities for the Riesz means of 
eigenvalues of order $\gamma {\ge} 3/4$ for a fourth order 
operator in arbitrary dimensions. We also consider some extensions 
to polyharmonic operators, and to systems of such operators, in 
dimensions greater than one. For the critical case $\gamma{=}1{-}1/(2l)$ in
dimension $d{=}1$ with $l {\geq} 2$ we prove the inequality 
$L^0_{l,\gamma,d}<L_{l,\gamma,d}$, which holds in contrast to current 
conjectures.
\end{abstract} 

\maketitle

\setcounter{section}{-1}
\section{Introduction} 
\subsection{Known facts}
Consider for $l \geq 1$ and $d \in \mathbb{N}$ the polyharmonic operator
$(-\Delta)^l + V$ in $L^2(\mathbb{R}^d)$, where $V$ is a real-valued function.
For suitable $V$ the negative spectrum of this operator is discrete. 
The Lieb-Thirring inequalities are estimates on the negative eigenvalues of
the form \footnote{Here and below we use the notion $2x_-:=|x|-x$ for the
  negative part of variables, functions, Hermitian matrices or self-adjoint
  operators.}
\begin{equation}
  \label{eq:PolyLiebTh}
  \tr ((-\Delta)^l + V)_-^\gamma \leq L_{l,\gamma,d}
\int_{\mathbb{R}^d}
V_-^{\gamma+\kappa}(x)\,dx, \quad V \in L^{\gamma+\kappa}(\mathbb{R}^d),
\end{equation}
which holds for certain $\gamma \geq 0$ with a constant
$L_{l,\gamma,d}$, depending only on $l, d$ and $\gamma$.
Here and in the following we use the abbreviations
\begin{equation*}
\kappa = \kappa(d,l) := \frac{d}{2 l}, \quad \nu = \nu(d,l):= 1 - 
\frac{d}{2 l}.
\end{equation*}
This type of inequalities was introduced by
Lieb and Thirring in \cite{LT}. They proved that \eqref{eq:PolyLiebTh} holds
in the case $l=1$ for all $\gamma > \max (0, \nu)$ 
with a finite constant $L_{l,\gamma,d}$. Their argument can easily be extended
to all $l \geq 1$.
On the other hand it is known that \eqref{eq:PolyLiebTh} fails for $\gamma = 0$
if $d = 2l$ and for $0 \leq \gamma < \nu$ if $d < 2l$.
In the critical case
$\gamma = 0, d > 2l $ the bound \eqref{eq:PolyLiebTh} exists and is for $l=1$ known as the
Cwikel-Lieb-Rosenblum inequality, see \cite{C,L1,R} and also
\cite{Con,LY}. The existence of $L_{l,\gamma,d}$ in the remaining critical
case $d < 2l, \gamma = \nu$ was verified by Netrusov and Weidl for \emph{integer}
values of $l$ in \cite{W1, W-N}.
Hence, the cases of existence for bounds of type \eqref{eq:PolyLiebTh} with $\gamma \geq 0$
are completely settled for integer $l$, while for non-integer
$l$ only the case $2l>d, \gamma = \nu$ is still open. 

For sufficiently regular potentials $V \in L^{\gamma + \kappa}(\mathbb{R}^d)$ the inequalities 
\eqref{eq:PolyLiebTh} are accompanied by the Weyl type asymptotic formula 
\begin{align}
  \label{Polyasyformu}
\notag
\lim_{\alpha\to+\infty}
\frac{1}
{\alpha^{\gamma+\kappa}}\,\tr ((-\Delta)^l+\alpha V)^\gamma_-&= 
\lim_{\alpha\to+\infty}\frac{1}{\alpha^{\gamma+\kappa}}
\iint_{\mathbb{R}^d\times\mathbb{R}^d}(|\xi|^{2l}+\alpha V)_-^\gamma\frac{dxd\xi}{(2\pi)^{d}}\\
&=L^{\mbox{\footnotesize\upshape cl}}_{l,\gamma,d}
\int_{\mathbb{R}^d} V_-^{\gamma+\kappa}\,dx\,,
\end{align}
where the so-called classical constant
$L^{\mbox{\footnotesize\upshape cl}}_{l,\gamma,d}$ 
is defined by
\begin{equation}\label{PolyLclass}
L^{\mbox{\footnotesize\upshape cl}}_{l,\gamma,d}=
\frac{\Gamma(\gamma+1)\Gamma(\kappa+1)}
{2^d\pi^{d/2}\Gamma(l \kappa+1)\Gamma(\kappa+\gamma+1)}\,,
\quad \gamma\geq 0\,.
\end{equation}
Formula \eqref{Polyasyformu} can be closed to \emph{all} potentials 
$V \in L^{\gamma + \kappa}(\mathbb{R}^d)$ if the bound \eqref{eq:PolyLiebTh} holds. 

Furthermore we consider the Lieb-Thirring constant for the ground state, that
is the smallest constant $L^0_{l,\gamma,d}$ which fulfils 
\begin{equation}\label{eq:defl0}
  \varkappa_0^\gamma \leq L^0_{l,\gamma,d} \int_{\mathbb{R}^d} V_-^{\gamma+\kappa}\,dx\,
\end{equation}
for all $V \in L^{\gamma + \kappa}(\mathbb{R}^d)$, where
$-\varkappa_0$ is the ground state of $(-\Delta)^l+ V$.
In the case $d < 2l$ with $\gamma = \nu$ the value of $L^0_{l,\nu,d}$ is given
by 
\begin{equation}
L^0_{l,\nu,d} = \frac{\pi\kappa}{\sin(\pi\kappa)} L^{\mbox{\footnotesize\upshape
    cl}}_{l,0,d} = \frac{1}{\nu} L^{\mbox{\footnotesize\upshape
    cl}}_{l,\nu,d},
\end{equation}
see \cite{W-N}. It is interesting to compare the value of the sharp constant  
$L_{l,\gamma,d}$ in \eqref{eq:PolyLiebTh} with the values of 
$L^{\mbox{\footnotesize\upshape cl}}_{l,\gamma,d}$ and $L^0_{l,\gamma,d}$.
In view of \eqref{Polyasyformu} and \eqref{eq:defl0} we immediately obtain that
\begin{equation}
\max(L^{\mbox{\footnotesize\upshape cl}}_{l,\gamma,d},L^0_{l,\gamma,d}) \leq L_{l,\gamma,d}
\end{equation}
for all $l, d$ and $\gamma$. One of the sparse results on exact values 
of $L_{l,\gamma,d}$ is due to Lieb and Thirring. In \cite{LT} they 
obtained for $d{=}l{=}1$, using the Buslaev-Faddeev-Zakharov trace 
formulae \cite{BF,FadZ}, that
\begin{equation}\label{L=Lclass1}
L_{l,\gamma,d} = L^{\mbox{\footnotesize\upshape cl}}_{l,\gamma,d}
\end{equation}
for $\gamma = 3/2 + n$ with $n \in \mathbb{N}_0$.
In \cite{A-L} Aizenman and Lieb found an argument, 
how to prove \eqref{L=Lclass1} in $d = 1$ for all $\gamma \geq 3/2$.
Applying a ``lifting'' argument with respect to dimension, Laptev and Weidl 
finally succeeded in \cite{L-W} to prove \eqref{L=Lclass1} for all $d \in
\mathbb{N}$ and $\gamma \geq 3/2$ in the case $l=1$.
In fact, their result is even more general, and is obtained for 
infinite-dimensional systems of Schr\"{o}dinger operators. 

However, in the case $l > 1$ no sharp constants are known, not even in 
dimension $d = 1$. In the paper \cite{Ost} an attempt was made to 
prove, that \eqref{L=Lclass1} holds for $d = 1, l = 2$ and $\gamma\geq 7/4$.
The constant appearing in \cite{Ost}, in the trace formula for the 
Riesz mean of order $7/4$, is precisely the classical, but whether 
the equality \eqref{L=Lclass1} holds true or not in that case is still open. 

The only other case where the sharp value of $L_{l,\gamma,d}$ is presently
known, is $d {=} l {=} 1$ with critical $\gamma = 1/2$, for which in \cite{HLT} it was proven by
Hundertmark, Lieb and Thomas that
\begin{equation}\label{eq:hlt}
L_{1,1/2,1}=L^0_{1,1/2,1} = 2\,L^{\mbox{\footnotesize\upshape cl}}_{1,1/2,1}=1/2\,.
\end{equation}
For the remaining cases the values of the Lieb-Thirring constants constitute an
interesting open problem. 

It shall be mentioned, that at least for the case $l=1$ there exists a
conjecture about the value of $L_{l,\gamma,d}$, which is due to Lieb and
Thirring \cite{LT}. The conjecture is, that for each dimension $d$ there
exists a unique $\gamma_c(d)$ so that 
\begin{align}
 &L_{1,\gamma,d} = L^{\mbox{\footnotesize\upshape cl}}_{1,\gamma,d}&
 &\mbox{for}\quad \gamma \geq \gamma_c(d)\quad\mbox{and}\label{eq:ltconjecture1}\\
 &L_{1,\gamma,d} = L^0_{1,\gamma,d}&  &\mbox{for}\quad \gamma  \leq
 \gamma_c(d).\label{eq:ltconjecture2}
\end{align}
Comparing this with the results above one sees, that
\eqref{eq:ltconjecture1} is proven to hold with $\gamma_c(d) \leq 3/2$ for all
$d \in \mathbb{N}$, where
\eqref{eq:ltconjecture2} is still open, but supported by \eqref{eq:hlt}.

%
%
%

\subsection{Main results of this paper.}
In section \ref{ltinequalitydimone} we follow the idea of \cite{HLTW} and
extend the argumentation in dimension one to the case $l>1$, which leads to
(non-sharp) inequalities for this case. See Theorem \ref{theorem1} for the
special case of the biharmonic operator $\partial^4+V$, and Theorem
\ref{theorem2} for the general case. Our results also apply to systems of
operators of the above kind, an issue raised in the paper \cite{LWR}, as well
as to non-integer $l$. We also discuss an extension of \cite{W1} to the case
$l=2$, see subsection \ref{sec:extnetweidl}. In section
\ref{estimatesfrombelow} we prove the inequality  
\begin{equation*}
  L^0_{l,\nu,1} < L_{l,\nu,1}
\end{equation*}
for integer $l\geq 2$, which holds in contrast to equality \eqref{eq:hlt}.
This answers a question posed in section 2.8 of [19] and, 
in particular, shows that the conjecture (0.10) does not apply 
to higher order operators. In section \ref{ltinequalitydimhigh} we lift the
results from 
section \ref{ltinequalitydimone} to higher dimensions, see especially Theorem
\ref{most-general}.

\section{Lieb-Thirring inequalities for Riesz means of eigenvalues 
for polyharmonic operators in dimension one}
\label{ltinequalitydimone}
\subsection{Notation and auxiliary material}\label{sec:notaux}
Let $\mathcal{G}$ be a separable Hilbert space with norm 
$||\cdot||_{\mathcal{G}}$ and scalar product
$\langle\cdot,\cdot\rangle_{\mathcal{G}}$. Further, let $\mathbf{0}_{\mathcal{G}}$ 
respectively $\mathbf{1}_{\mathcal{G}}$ be the zero respectively identity 
operator on $\mathcal{G}$, and $\mathcal{B}\left(\mathcal{G}\right)$ be
the Banach space of bounded operators on $\mathcal{G}$. The Hilbert space 
$\mathcal{H} := L^{2}\left(\mathbb{R}^{d},\mathcal{G}\right)$ is the space
of all measurable functions $u : \mathbb{R}^{d} \rightarrow \mathcal{G}$
such that 
\begin{equation*}
||u||_{\mathcal{H}}^{2} \,:= \int_{\mathbb{R}^{d}} ||u(x)||_{\mathcal{G}}^{2}
  \, dx < \infty.
\end{equation*}   
The scalar product in $\mathcal{H}$ is given by 
\begin{equation*}
\langle u,v\rangle_{\mathcal{H}} \,:= \int_{\mathbb{R}^{d}} 
\langle u(x),v(x)\rangle_{\mathcal{G}} \, dx, \quad \mbox{for } u,v \in \mathcal{H}.
\end{equation*}
The space $L^{2}\left(\mathbb{R}^{d}, \mathcal{G}\right)$ is
naturally isomorphic to 
$L^{2}\left(\mathbb{R}^{d}\right) \otimes \mathcal{G}$,
and we will make no distinction between them. We shall denote by 
$\Phi$ the Fourier transform unitary on $L^{2}\left(\mathbb{R}^{d}\right)$.
For simplicity of notation, whenever 
$u \in L^{2}\left(\mathbb{R}^{d}, \mathcal{G}\right)$ we further let
$\hat{u} := (\Phi \otimes \mathbf{1}_{\mathcal{G}})\,u$. 
The Sobolev space $H^{l}\left(\mathbb{R}^{d},\mathcal{G}\right)$,
for $l > 0$, is the subset of $L^{2}\left(\mathbb{R}^{d}, \mathcal{G}\right)$ 
defined by 
\begin{equation*}
H^{l}\left(\mathbb{R}^{d},\mathcal{G}\right) := 
\left\{ u \in L^{2}(\mathbb{R}^{d}, \mathcal{G}) : 
\left(1 + |\xi|^2\right)^{l/2}\,\hat{u}(\xi) 
\in L^{2}\left(\mathbb{R}^{d},\mathcal{G}\right)\right\}. 
\end{equation*}
The space $H^{l}\left(\mathbb{R}^{d},\mathcal{G}\right)$, 
equipped with the scalar product
\begin{equation*}
\langle u, v\rangle_{H^{l}\left(\mathbb{R}^{d},\mathcal{G}\right)} \, :=
\int_{\mathbb{R}^{d}} \left(1 + |\xi|^2\right)^{l}\,
\langle\hat{u}(\xi), \hat{v}(\xi)\rangle_{\mathcal{G}}\,d\xi,
\end{equation*}
is a Hilbert space.
As in the scalar case $\mathcal{G} = \mathbb{C}$ one sees that 
if $l \in \mathbb{N}$, then
\begin{equation*}
H^{l}\left(\mathbb{R}^{d},\mathcal{G}\right) = 
\left\{ u \in L^{2}(\mathbb{R}^{d}, \mathcal{G}) : 
\partial^{\alpha}\,u \in L^{2}\left(\mathbb{R}^{d},\mathcal{G}\right),
\quad |\alpha| \leq l \right\}. 
\end{equation*} 
Obviously, for $l > 0$, the quadratic form 
\begin{equation*}
h[u,u] \, := \int_{\mathbb{R}^{d}} |\xi|^{2l}\,
||\hat{u}(\xi)||^{2}_{\mathcal{G}}\,d\xi
\end{equation*}
is semibounded from below and closed on the form-domain
$H^{l}\left(\mathbb{R}^{d},\mathcal{G}\right) \subset 
L^{2}\left(\mathbb{R}^{d},\mathcal{G}\right)$. It is associated with the 
self-adjoint operator $\left(- \Delta\right)^{l} \otimes 
\mathbf{1}_{\mathcal{G}}$ on 
$H^{2l}\left(\mathbb{R}^{d},\mathcal{G}\right)$.

Let $V : \mathbb{R}^{d} \rightarrow \mathcal{B}\left(\mathcal{G}\right)$ be
an operator-valued function, for which $V(x) = (V(x))^{*}$ for a.e. 
$x \in \mathbb{R}^{d}$, satisfying:
\begin{equation}
  \label{Vconditions}
||V(\cdot)||_{\mathcal{B}\left(\mathcal{G}\right)} \in
  L^{p}\left(\mathbb{R}^{d}\right)
\end{equation}
with some finite $p$ with 
\begin{equation*}
\begin{array}{ll}
p \geq 1 \quad &\mbox{if } d < 2l,\\
p > 1 \quad &\mbox{if } d = 2l,\\
p \geq d/2l \quad &\mbox{if } d > 2l.  
\end{array}
\end{equation*}
Then the form 
\begin{equation*}
\nu[u,u] := \int_{\mathbb{R}^{d}} \langle V\,u , u\rangle_{\mathcal{G}} \, dx
\end{equation*}
is well-defined on $H^{l}\left(\mathbb{R}^{d},\mathcal{G}\right)$ and 
\begin{equation}
  \label{boundedness}
|\nu[u,u]| \leq C \left(\int_{\mathbb{R}^{d}} 
||V||^{p}_{\mathcal{B}\left(\mathcal{G}\right)} \,
dx\right)^{1/p} ||u||_{H^{l}\left(\mathbb{R}^{d},\mathcal{G}\right)}^{2}.
\end{equation}
This follows from analogs of the standard Sobolev imbedding theorems 
which hold in the scalar case. For instance, in case $d > 2l$, this 
follows from H\"{o}lder's inequality and the imbedding 
$H^{l}\left(\mathbb{R}^{d},\mathcal{G}\right) \hookrightarrow 
L^{q}\left(\mathbb{R}^{d}, \mathcal{G}\right)$, $q \leq q^{*} := 
\frac{2d}{d - 2l}$.
Moreover, for all $\epsilon > 0$ there exists a constant $C(\epsilon,
V)$ such that 
\begin{equation}
|\nu[u,u]| \leq \epsilon h[u,u] + C(\epsilon, V) \int_{\mathbb{R}^{d}} 
||u||_{\mathcal{G}}^{2} \, dx, 
\quad u \in H^{l}\left(\mathbb{R}^{d},\mathcal{G}\right). 
\end{equation}
This is also a version of the corresponding inequality which is well-known
in the scalar case, when $\mathcal{G} = \mathbb{C}$.
It follows that the form 
\begin{equation*}
h[u,u] + \nu[u,u]
\end{equation*}
is semibounded from below and closed on 
$H^{l}\left(\mathbb{R}^{d},\mathcal{G}\right)$. It induces a
self-adjoint operator
\begin{equation}
  \label{OperatorQ}
Q := \left(- \Delta\right)^{l} \otimes \mathbf{1}_{\mathcal{G}} + V
\end{equation}
in $\mathcal{H} = L^{2}\left(\mathbb{R}^{d},\mathcal{G}\right)$.\\
More precise conditions guaranteeing $V$ to be a weak Hardy weight, 
stated in terms of capacities, are given in \cite{Mz}.

If $V$ satisfies the condition \eqref{Vconditions} and
if $V(x) \in S_{\infty}\left(\mathcal{G}\right)$ for a.e. 
$x \in \mathbb{R}^{d}$, the negative spectrum of the operator $Q$ is 
discrete and might accumulate only to $0$. In other words, the 
operator $Q_{-}$ is compact in 
$\mathcal{H} = L^{2}\left(\mathbb{R}^{d},\mathcal{G}\right)$. 
This can be proven as follows. We clearly may assume $V \leq 0$, by 
the minimax principle, and put $W := \sqrt{-V}$.  
By the Birman-Schwinger principle, for $\varkappa > 0$, the number 
$N_{-}\left(-\varkappa, Q\right)$ of eigenvalues of $Q$ less than
$-\varkappa$ equals the number $N_{+}\left(1, B_{W}\left(\varkappa\right)\right)$ 
of eigenvalues greater than $1$ of the Birman-Schwinger operator
\begin{equation*}
B_{W}(\varkappa) := W\left( \left(-\Delta\right)^{l} \otimes 
\mathbf{1}_{\mathcal{G}} + \varkappa\right)^{-1} W 
\end{equation*}
on $L^{2}\left(\mathbb{R}^d, \mathcal{G}\right)$. One sees that 
$B_{W}(\varkappa) = S_{W}\,S_{W}^{*}$, where
\begin{equation*}
S_{W} := W \left(\Phi^{*} \otimes \mathbf{1}_{\mathcal{G}}\right) 
\left(|\xi|^{2l} + \varkappa\right)^{-1/2}.
\end{equation*}
Thus the claim follows by compactness of $S_{W}$ on 
$L^{2}\left(\mathbb{R}^d, \mathcal{G}\right)$.

\subsection{Estimates of Riesz means for the biharmonic operator in d = 1}
\label{fourthorderfullspace}
In this section we obtain the following:
\begin{theorem}
  \label{theorem1}
Let $V : \mathbb{R} \rightarrow \mathcal{B}\left(\mathcal{G}\right)$ be
an operator-valued function satisfying $V(x) = (V(x))^{*}$ and 
$V(x) \in S_{1}\left(\mathcal{G}\right)$ for a.e. $x \in \mathbb{R}$ and
such that $\tr V_{-}(\cdot) \in L^{1}\left(\mathbb{R}\right)$. 
Then the following inequality holds true:
\begin{equation}
  \label{betterconst}
  \tr \left(\partial^4 \otimes \mathbf{1}_{\mathcal{G}} +
  V\right)_{-}^{3/4} \leq
  \frac{3^{3/4}}{4} \int_\mathbb{R} \tr V_{-}(x) \, dx.
\end{equation}
\end{theorem}
The original proof of the analog of Theorem \ref{theorem1} for 
Schr\"{o}dinger operators \mbox{$-\partial^2 + V$} was given in the
paper by Hundertmark, Lieb and Thomas \cite{HLT}. Here we 
follow closely the argument in the proof of the same statement given 
by Hundertmark, Laptev and Weidl in \cite{HLTW}.
   
For the proof of the theorem we need to introduce some auxiliary
results on the notion of ``majorization''. Let $A$ be a compact
operator on a separable Hilbert space $H$. 
Let us denote 
\begin{equation}
||A||_{n} := \sum_{j=1}^{n} \sqrt{\lambda_{j}(A^{*}A)},
\end{equation}
where $(\lambda_j(A^{*}A))_j$ is the sequence of the eigenvalues of $A^{*}A$ in
 non-increasing order according to their multiplicities.
 Then by Ky-Fan's inequality (see for instance \cite{G-K}) the functionals $||\cdot||_{n}$ are 
norms on $S_{\infty}(H)$, and for any unitary operator 
$\mathcal{U}$ in $H$ we have
\begin{equation*}
||\, \mathcal{U}^{*} A \, \mathcal{U}||_{n} = ||A||_{n}.
\end{equation*}  
We shall need the following definition and lemma, which were stated in \cite{HLTW}.
\begin{definition} Let $A$, $B$ be any two compact operators on $H$. We
say that $A$ majorizes $B$, written $B \prec A$, if 
\begin{equation*}
||B||_{n} \leq ||A||_{n} \quad \mbox{for all }\quad n \in \mathbb{N}.
\end{equation*}
\end{definition}
\begin{lemma}
  \label{majorizationlemma}
Let $A$ be a non-negative compact operator on $H$, 
$\{\mathcal{U}(\omega)\}_{\omega \in \Omega}$ a weakly measurable family of unitary
operators on $H$, and $\mu$ a probability measure on $\Omega$.  Then 
the operator 
\begin{equation*}
B := \int_{\Omega} \mathcal{U}^{*}(\omega) A \, \mathcal{U}(\omega) \, 
d\mu(\omega)
\end{equation*}
is majorized by the operator $A$.
\end{lemma}
\begin{proof}[\textbf{Proof.}]  This follows immediately from Ky-Fan's inequality:
\begin{equation*}
||B||_{n} \leq \int_{\Omega} ||\,\mathcal{U}^{*}(\omega) A
\mathcal{U}(\omega) ||_{n} \, d\mu (\omega) = \mu(\Omega) || A||_{n} 
= || A||_{n}, \quad n \in \mathbb{N}.  
\end{equation*} 
\end{proof}

By the minimax principle we may assume $V$ non-positive and put 
$W := \sqrt{-V}$. We shall have use of the following family of
operators on $\mathcal{H} = L^{2}(\mathbb{R},\mathcal{G})$:
\begin{alignat*}{4}
\mathcal{L}_{\epsilon} &:= W \left[\epsilon^{3} 
\left(\partial^4 + \epsilon^{4}\right)^{-1} \otimes 
\mathbf{1}_{\mathcal{G}}\right] W,\\ 
\tilde{\mathcal{L}}_{\epsilon} &:= W \left[ a \epsilon 
\left(-\partial^2 + \epsilon^{2} b\right)^{-1} \otimes 
\mathbf{1}_{\mathcal{G}}\right] W,
\end{alignat*}
for $0 < \epsilon < \infty$. Furthermore, let us define 
$\tilde{\mathcal{L}}_{0} := A$, where $A$ is the non-negative
compact operator having integral-kernel 
$A(x,y) := \frac{a}{2 \sqrt{b}} \, W(x) \, W(y)$. 
The positive constants $a$ and $b$ will be
specified later. The following result is almost identical to a lemma 
proven in \cite{HLTW}.
\begin{lemma}
  \label{Lmajorization}
The operator $\tilde{\mathcal{L}}_{\epsilon}$ is majorized by 
$\tilde{\mathcal{L}}_{\epsilon'}$, 
\begin{equation*}
\tilde{\mathcal{L}}_{\epsilon} \prec \tilde{\mathcal{L}}_{\epsilon'}
\end{equation*}
for all $0 \leq \epsilon' < \epsilon$.
\end{lemma} 
\begin{proof}[\textbf{Proof.}]  We shall use the majorization Lemma
\ref{majorizationlemma}.
Introduce a family of probability measures $\mu_{\epsilon}$ on $\mathbb{R}$ by
$\mu_{0} := \delta_{0}$, the Dirac measure, and  
\begin{equation*}
\frac{d\mu_{\epsilon}}{d\xi} = \frac{\epsilon \sqrt{b}}{\pi} \,
\frac{1}{\xi^2 + \epsilon^2 b}=:g_{\epsilon}(\xi), \quad \epsilon > 0,
\end{equation*}
where $d\xi$ denotes the Lebesgue measure.  
Furthermore, let $\{\mathcal{U}(\xi)\}_{\xi \in \mathbb{R}}$ be the unitary
multiplication operators in $L^{2}(\mathbb{R},\mathcal{G})$ defined by 
$(\mathcal{U}(\xi)u)(x) = e^{-i \xi x} \, u(x)$. We then see that 
\begin{equation}
  \label{Fourierrepresentation}
\tilde{\mathcal{L}}_{\epsilon} = \int_{\mathbb{R}} \mathcal{U}^{*}(\xi)
A \, \mathcal{U}(\xi) \, d\mu_{\epsilon},
\end{equation} 
for any $0 \leq \epsilon < \infty$. It follows from Lemma 
\ref{majorizationlemma} and \eqref{Fourierrepresentation} that 
$\tilde{\mathcal{L}}_{\epsilon} \prec \tilde{\mathcal{L}}_{0}$.
Since the Fourier transform of $g_{\epsilon}$ is 
given by 
\begin{equation*}
\Phi g_{\epsilon}(\xi) = \frac{1}{\sqrt{2 \pi}} \, e^{- \epsilon \sqrt{b}
  \, |\xi|}, \quad \mbox{for } \epsilon > 0,
\end{equation*}
it follows that for any $0 < \epsilon' < \epsilon$
\begin{equation}
  \label{convolutionidentity}
g_{\epsilon} = g_{\epsilon'} \, * \, g_{\epsilon - \epsilon'}.
\end{equation}
Using the relation \eqref{convolutionidentity} in 
\eqref{Fourierrepresentation} as well as the group property of the
unitary operators $\mathcal{U}(\xi)$ it is now seen that 
\begin{equation}
\tilde{\mathcal{L}}_{\epsilon} = \int_{\mathbb{R}} \mathcal{U}^{*}(\eta)
\tilde{\mathcal{L}}_{\epsilon'} \, \mathcal{U}(\eta)  
g_{\epsilon - \epsilon'}(\eta) \, d\eta 
\prec \tilde{\mathcal{L}}_{\epsilon'},
\end{equation}   
where the last subordination follows from Lemma \ref{majorizationlemma}.
This completes the proof. 
\end{proof}

We are now in the position of proving the above theorem.\\
\begin{proof}[\textbf{Proof of Theorem \ref{theorem1}.}] First note that if we put $a := 
\frac{b + \sqrt{b^2 + 1}}{2}$, then for any $b > 0$
\begin{equation}
  \label{Linequality}
\mathcal{L}_{\epsilon} \leq \tilde{\mathcal{L}}_{\epsilon}.
\end{equation}   
This follows immediately from the computation 
\begin{alignat*}{4}
\langle\mathcal{L}_{\epsilon} \, u, u\rangle_{\mathcal{H}} &= \int_{\mathbb{R}} 
\frac{\epsilon^{3}}{\xi^4 + \epsilon^4} 
||(\Phi \otimes \mathbf{1}_{\mathcal{G}}) \, W \, u \, (\xi)
||_{\mathcal{G}}^2 \, d\xi \\
&\leq \int_{\mathbb{R}} \frac{a \epsilon}{\xi^2 + \epsilon^2 b}
||(\Phi \otimes \mathbf{1}_{\mathcal{G}}) \, W \, u \, (\xi)
||_{\mathcal{G}}^2 \, d\xi \\
&= \, \langle\tilde{\mathcal{L}}_{\epsilon} \, u, u\rangle_{\mathcal{H}}, 
\quad u \in L^{2}\left(\mathbb{R},\mathcal{G}\right),  
\end{alignat*}
which holds in view of the scalar inequality  
\begin{equation*}
\frac{\epsilon^{2}}{\xi^4 + \epsilon^4} \leq \frac{a}{\xi^2 + \epsilon^2 b}.
\end{equation*} 
Here $\Phi$ denotes the Fourier transform on $L^{2}(\mathbb{R})$. 
For $E > 0 $ let us define
\begin{equation}
\mathcal{K}_{E} := \frac{1}{E^{3/4}} \mathcal{L}_{E^{1/4}}
= W \left[(\partial^4 + E)^{-1} \otimes \mathbf{1}_{\mathcal{G}} \right] W.
\end{equation}
Denote by $(-E_{j})_j$ the negative eigenvalues of the operator
$\partial^4 + V$, and by $(\lambda_{j}(T))_j$ the eigenvalues of a
non-negative compact operator $T$, enumerated according to their
multiplicities in non-decreasing respectively non-increasing order. By the
Birman-Schwinger principle  
\begin{equation}
  \label{Birman-Schwinger}
1 = \lambda_{j} (\mathcal{K}_{E_{j}}).
\end{equation} 
Multiplying the identity \eqref{Birman-Schwinger} by 
$E_{j}^{3/4}$ and summing over $j$ we get from \eqref{Linequality} 
and the minimax principle
\begin{equation}
  \label{Birman-Schwinger1}
\sum_{j} E_{j}^{3/4} = \sum_{j} \lambda_{j} 
( \mathcal{L}_{E_{j}^{1/4}} ) \leq \sum_{j} \lambda_{j} 
( \tilde{\mathcal{L}}_{E_{j}^{1/4}} ).
\end{equation}
The interesting point now is that although the trace of 
$\tilde{\mathcal{L}}_{\epsilon}$ is independent of $\epsilon$, by
Lemma \ref{Lmajorization} the partial traces 
$\sum_{j \leq n} \lambda_{j}(\tilde{\mathcal{L}}_{\epsilon} )$ are
monotone decreasing in $\epsilon$ for any $n \in \mathbb{N}$. It follows 
that 
\begin{equation}
  \label{Birman-Schwinger2}
\sum_{j \leq n} \lambda_{j} ( \tilde{\mathcal{L}}_{E_{j}^{1/4}} ) \leq  
\sum_{j \leq n} \lambda_{j} ( \tilde{\mathcal{L}}_{E_{n}^{1/4}} ) \leq
\sum_{j \leq n} \lambda_{j} ( \tilde{\mathcal{L}}_{0} ), \quad 
\mbox{ for all } n \in \mathbb{N}. 
\end{equation}
The first inequality above follows from this monotonicity by induction 
over $n \in \mathbb{N}$, the second from the monotonicity directly. Combining 
\eqref{Birman-Schwinger1} and \eqref{Birman-Schwinger2} gives 
\begin{equation*}
\sum_{j} E_{j}^{3/4} \leq \tr \tilde{\mathcal{L}}_{0},
\end{equation*}
where
\begin{equation}\label{eq:minib}
\tr \tilde{\mathcal{L}}_{0} = \frac{a}{2 \sqrt{b}} \int \tr V_{-}(x)\, dx
= \frac{b + \sqrt{b^2 + 1}}{4 \sqrt{b}} \int \tr V_{-}(x)\, dx. 
\end{equation}
Minimizing the right hand side of \eqref{eq:minib} with respect to $b$
leads to the choice $b := 1/\sqrt{3}$, and an
evaluation of the expression completes the proof. 
\end{proof}

Applying the Aizenman-Lieb argument from \cite{A-L}, we obtain the 
following corollary: 
\begin{corollary}
  \label{Biharmoniccorollary}
Let $V : \mathbb{R} \rightarrow \mathcal{B}\left(\mathcal{G}\right)$ be 
an operator-valued function satisfying $V(x) = (V(x))^{*}$ and 
$V(x) \in S_{1}\left(\mathcal{G}\right)$ for a.e. $x \in \mathbb{R}$ and
such that $\tr V_{-}(\cdot) \in L^{\gamma+\frac{1}{4}}\left(\mathbb{R}\right)$, 
for some $\gamma \geq 3/4$.
Then the following inequality holds true:
\begin{equation}
\tr \left(\partial^4 \otimes \mathbf{1}_{\mathcal{G}} + V\right)_{-}^{\gamma} \leq
\frac{4}{3^{1/4} \sqrt{2}} \, L^{\mbox{\footnotesize\upshape cl}}_{2,\gamma,1}
\int_{\mathbb{R}} \tr \left(V_{-}(x)\right)^{\gamma + \frac{1}{4}} \, dx.
\end{equation}
\end{corollary}
\noindent
\textbf{Remark.} A numerical calculation yields $\frac{4}{3^{1/4} \sqrt{2}} \approx 2.149$.
\begin{proof}[\textbf{Proof.}]  
First note that for $\gamma > 3/4$
\begin{equation*}
\int_{0}^{\infty} t^{\gamma - \frac{7}{4}} \left(t +
\lambda\right)_{-}^{3/4} \, dt = \lambda_{-}^{\gamma} \,
B\left(\gamma - \frac{3}{4}, \frac{7}{4}\right),
\end{equation*} 
where $B(x,y) = \frac{\Gamma(x) \Gamma(y)}{\Gamma(x+y)}$ is the
Beta-function. Let $E_{Q}$ be the spectral measure associated with 
the self-adjoint operator $Q = \partial^{4} \otimes \mathbf{1}_{\mathcal{G}} + V$
and denote by $(-\mu_{j}(x))_j$ the negative eigenvalues of the operator 
$V(x)$. Since 
\begin{equation*}
\tr Q_{-}^{\gamma} = \tr \, \int_{\mathbb{R}} \lambda_{-}^{\gamma} \, 
dE_{Q}(\lambda) 
\end{equation*}
we obtain 
\begin{alignat*}{4}
&B\left(\gamma - \frac{3}{4}, \frac{7}{4}\right) \, \tr Q_{-}^{\gamma} = 
\tr \left\{\int_{\mathbb{R}} dE_{Q}(\lambda) \int_{0}^{\infty} 
t^{\gamma - \frac{7}{4}} \left(t + \lambda\right)_{-}^{3/4} \, dt\right\} \\
&= \tr \left\{\int_{0}^{\infty} dt \,t^{\gamma - \frac{7}{4}} \int_{\mathbb{R}}
dE_{Q}(\lambda) (t + \lambda)_{-}^{3/4} \right\} 
= \int_{0}^{\infty} t^{\gamma - \frac{7}{4}} \, \tr (t + Q)_{-}^{3/4} \,dt \\
&\leq \frac{3^{3/4}}{4} \, \int_{0}^{\infty} t^{\gamma - \frac{7}{4}}
\int_{\mathbb{R}} \tr \left(t + V(x)\right)_{-} dx \, dt \\
&= \frac{3^{3/4}}{4} \, \int_{\mathbb{R}} \,dx \int_{0}^{\infty} 
t^{\gamma - \frac{7}{4}} \sum_{j=1}^{\infty} \left(t -
\mu_{j}(x)\right)_{-} \, dt \\
&= \frac{3^{3/4}}{4} \, B\left(\gamma - \frac{3}{4},2\right) 
\int_{\mathbb{R}} \tr \left(V_{-}(x)\right)^{\gamma + \frac{1}{4}} \, dx.
\end{alignat*}
It follows that 
\begin{equation*}
\tr \left(\partial^4 \otimes \mathbf{1}_{\mathcal{G}} + V\right)_{-}^{\gamma}
\leq \frac{3^{3/4}}{4} \,
\frac{B\left(\gamma - \frac{3}{4},2\right)}{B\left(\gamma -
  \frac{3}{4}, \frac{7}{4}\right)} 
\int_{\mathbb{R}} \tr \left(V_{-}(x)\right)^{\gamma + \frac{1}{4}} \, dx.
\end{equation*}
The proof ends by noting that 
\begin{alignat*}{4}
L^{\mbox{\footnotesize\upshape cl}}_{2,\gamma,1} &= 
\frac{\Gamma(\gamma + 1) \, \Gamma(5/4)}{2 \sqrt{\pi} \, \Gamma(3/2) 
\, \Gamma(\gamma + 5/4)} = \frac{\Gamma(7/4) \, \Gamma(5/4)}{2 \sqrt{\pi} \, \Gamma(3/2) 
\, \Gamma(2)} \cdot \frac{\Gamma(\gamma + 1) \, \Gamma(2)}{\Gamma(\gamma + 5/4) \, \Gamma(7/4)} \\
&=  \frac{3 \sqrt{2}}{16} \cdot \frac{\Gamma(\gamma + 1) \, 
\Gamma(2)}{\Gamma(\gamma + 5/4) \, \Gamma(7/4)} = \frac{3 \sqrt{2}}{16} 
\cdot \frac{B\left(\gamma - \frac{3}{4},2\right)}
{B\left(\gamma - \frac{3}{4}, \frac{7}{4} \right)}.
\end{alignat*} 
\end{proof} 

\subsection{Results by the method from Netrusov and Weidl \cite{W1,W-N}}
\label{sec:extnetweidl}
%
%
%
%
%
%
%

This method is based on a special Neumann-bracketing technique which
together with the Birman-Schwinger principle leads
to rather implicit bounds for the Lieb-Thirring constants in the case
$2l > d$ with critical $\gamma = 1-\frac{d}{2l}$. In \cite{diplfoerster} a detailed
analysis of the case $l=2, d=1$ was done, which yields for the corresponding
Lieb-Thirring constant $L_{2,3/4,1}$ the estimate
\begin{equation}\label{eq:second4ordestimate}
  L_{2,3/4,1} < 2.129.
\end{equation}
This estimate is much worse than \eqref{betterconst}. Its value is
mainly, that it is also an upper estimate on the Lieb-Thirring constant
$L^+_{2,\frac34,1}$ for the operator $\partial^4 + V$
in $L^2\big((0,\infty)\big)$ with \emph{Neumann} conditions in zero. Notice, that
Neumann conditions mean here, that the second and third derivative vanish at
zero. 

We remark furthermore that the unique negative eigenvalue $-\varkappa_+$ of the
Neumann operator $\partial^4 - \delta_0$ in $L^2\big((0,\infty)\big)$,
associated with the quadratic form
\begin{equation*}
  h^+[u,u] := \| \partial^2 u\|^2 - |u(0)|^2,\quad u \in
  H^2\big((0,\infty)\big),
\end{equation*}
fulfils $$\varkappa_+^{3/4} = \sqrt2.$$ So for the
half space problem the inequality  
\begin{equation}\label{eq:halfspaceestimate}
\sqrt{2} \leq L^+_{2,\frac34,1} < 2.129.
\end{equation}
holds. 
\subsection{Estimates of Riesz means for polyharmonic operators in d = 1}
It is not difficult to adapt the proof of Theorem \ref{theorem1} to
polyharmonic operators of the form
\begin{equation}
\left(-\partial^2\right)^{l} \otimes \mathbf{1}_{\mathcal{G}} + V, 
\quad l > 1.
\end{equation}
We obtain the following theorem:
\begin{theorem}\label{theorem2}
Let $V : \mathbb{R} \rightarrow \mathcal{B}\left(\mathcal{G}\right)$ be 
an operator-valued function satisfying $V(x) = (V(x))^{*}$ and 
$V(x) \in S_{1}\left(\mathcal{G}\right)$ for a.e. $x \in \mathbb{R}$ and
such that $\tr V_{-}(\cdot) \in L^{1}\left(\mathbb{R}\right)$.
Then the Riesz mean for the critical power $\nu = 1 - 1/2l$ of the
polyharmonic operator $\left(-\partial^2\right)^{l} + V$, $l > 1$,
satisfies the bound
\begin{equation}
\tr \left(\left(-\partial^2\right)^{l} \otimes \mathbf{1}_{\mathcal{G}} 
+ V\right)_{-}^{1-1/2 l} \leq c_{l} \int \tr V_{-}(x) \, dx.
\end{equation}
Here the constant $c_{l}$ is defined as 
\begin{equation}
c_{l} := \frac{1}{2 l} \zeta_{l}^{l - 1},
\end{equation}
where $\zeta_{l}$ is the unique positive root of the equation
\begin{equation}
\left(l-1\right) + l z - z^{l} = 0. 
\end{equation}
\end{theorem}
\begin{proof}[\textbf{Proof.}]  Define the following family of operators on 
$L^{2}\left(\mathbb{R}, \mathcal{G}\right)$:
\begin{alignat*}{4}
\mathcal{L}_{\epsilon} &:= W \left[\epsilon^{2 l - 1} 
\left( \left(-\partial^2\right)^{l} + \epsilon^{2 l}\right)^{-1} \otimes 
\mathbf{1}_{\mathcal{G}}\right] W,\\ 
\tilde{\mathcal{L}}_{\epsilon} &:= \tilde{c}_{l} \, W \left[\epsilon 
\left(-\partial^2 + \epsilon^{2}\right)^{-1} \otimes \mathbf{1}_{\mathcal{G}}
\right] W,
\end{alignat*}
for $0 < \epsilon < \infty$. Here the constant $\tilde{c}_{l}$ is
defined as
\begin{equation}
\tilde{c}_{l} : = \sup_{x,y>0} \frac{y^{2l - 2} \left(x^{2} +
y^{2}\right)}{x^{2l} + y^{2l}}.
\end{equation}
As before we may assume $V$ non-positive and put $W := \sqrt{-V}$. 
In view of the scalar inequality 
\begin{equation}
  \label{scalarinequality}
\frac{\epsilon^{2l - 1}}{\xi^{2l} + \epsilon^{2l}} \leq  
\tilde{c}_{l} \frac{\epsilon}{\xi^2 + \epsilon^2},
\end{equation}
we then see that 
\begin{equation*}
\mathcal{L}_{\epsilon} \leq \tilde{\mathcal{L}}_{\epsilon}.
\end{equation*}
We may therefore proceed similarly as in the proof above, to obtain
\begin{equation*}
\tr \left(\left(-\partial^2\right)^{l} \otimes \mathbf{1}_{\mathcal{G}} 
+ V\right)_{-}^{1-1/2 l} \leq \frac{1}{2} \, \tilde{c}_{l} 
\int \tr V_{-} \, dx.
\end{equation*}
It remains only to prove that 
\begin{equation}
\tilde{c}_{l} = \frac{1}{l} \zeta_{l}^{l - 1}. 
\end{equation}
But a glance at the function $f_{l}$ defined by 
\begin{equation}
f_{l}(x,y) := \frac{y^{2l - 2} \left(x^{2} +
y^{2}\right)}{x^{2l} + y^{2l}}, \quad x,y > 0,
\end{equation}
reveals that it attains constant values on the lines $y = \rho x$, $\rho > 0$. 
In fact,
\begin{equation}
f_{l}(x,\rho x) = \frac{\rho^{2l - 2} + \rho^{2l}}{1 + \rho^{2l}} =: g(\rho), 
\quad \rho > 0. 
\end{equation} 
A simple computation gives that 
\begin{equation} 
g'(\rho) = \frac{2 \rho^{2l -3}}{(1 + \rho^{2l})^{2}} \left(\left(l-1\right) -
\rho^{2l} + l \rho^{2}\right).
\end{equation}
We see that $g$ attains its maximal value in the critical points
$\rho_{l}$ given as the solution of 
\begin{equation*}
(l - 1) + l \rho^{2} - \rho^{2l} = 0.   
\end{equation*}
The maximal value of the function g attained at the critical points
$\rho_{l}$ is seen to be
\begin{equation*}
g(\rho_{l}) = \frac{1}{l} \rho_{l}^{2(l-1)}.
\end{equation*}  
The theorem follows by putting $\zeta_{l} := \rho_{l}^{2}$. 
\end{proof}  
The Aizenman-Lieb \cite{A-L} argument gives:
\begin{corollary}
  \label{Polyharmoniccorollary}
Let $V : \mathbb{R} \rightarrow \mathcal{B}\left(\mathcal{G}\right)$ be 
an operator-valued function satisfying $V(x) = (V(x))^{*}$ and 
$V(x) \in S_{1}\left(\mathcal{G}\right)$ for a.e. $x \in \mathbb{R}$ and
such that $\tr V_{-}(\cdot) \in
L^{\gamma+\frac{1}{2l}}\left(\mathbb{R}\right)$, for some $\gamma \geq
1 - 1/2l$, $l > 1$.
Then the following inequality holds true:
\begin{equation}
\tr \left((-\partial^2)^{l} \otimes \mathbf{1}_{\mathcal{G}} + V\right)_{-}^{\gamma} 
\leq
\frac{c_{l}}{L^{\mbox{\footnotesize\upshape cl}}_{l,1-1/2l,1}} 
\, L^{\mbox{\footnotesize\upshape cl}}_{l,\gamma,1}
\int_{\mathbb{R}} \tr \left(V_{-}(x)\right)^{\gamma + \frac{1}{2l}} \, dx.
\end{equation}
Here the constant $c_{l}$ is the same as in the above theorem.
\end{corollary}

\begin{proof}[\textbf{Proof.}]  The proof is almost identical to that for the biharmonic 
operator. We put $\nu:= 1 - 1/2l$ and note that 
\begin{equation*}
\int_{0}^{\infty} t^{\gamma - (1+\nu)} \left(t +
\lambda\right)_{-}^{\nu} \, dt = \lambda_{-}^{\gamma} \,
B\left(\gamma - \nu,1 + \nu\right).
\end{equation*}
We use this similarly as above to verify 
\begin{equation*}
\tr \left((-\partial^{2})^l \otimes \mathbf{1}_{\mathcal{G}} + V\right)_{-}^{\gamma}
\leq c_{l} \,
\frac{B\left(\gamma - \nu,2\right)}{B\left(\gamma - \nu,1 + \nu\right)} 
\int_{\mathbb{R}} \tr \left(V_{-}(x)\right)^{\gamma + \frac{1}{2l}} \, dx.
\end{equation*}
Finally we verify that 
\begin{equation*}
L^{\mbox{\footnotesize\upshape cl}}_{l,\gamma,1} = 
 L^{\mbox{\footnotesize\upshape cl}}_{l,\nu,1} \cdot 
\frac{B\left(\gamma - \nu,2\right)}{B\left(\gamma - \nu,1 + \nu\right)}.
\end{equation*} 
\end{proof} 

\section{Estimates of Lieb-Thirring constants from below in dimension one}
\label{estimatesfrombelow}
In this section we prove the following result, where the emphasis is on the
\emph{strict} inequality \eqref{eq:lbiggerl0}.
\begin{theorem}\label{theo:lbiggerl0}
  For $l \in \mathbb{N}$ with $l \geq 2$ and $\nu = 1 - \frac{1}{2l}$ the inequality
  \begin{equation}\label{eq:lbiggerl0}
    L^0_{l,\nu,1} < L_{l,\nu,1}
  \end{equation}
  holds true.
\end{theorem}

We point out the difference to the case $l=1$, where
one has equality in \eqref{eq:lbiggerl0}. This difference originates in the fact, that the
eigenfunction corresponding to the ground state of
$(-\partial^2)^l - \delta_0$ has no zeros for $l=1$, whereas it has zeros for
all $l \geq 2$. The idea of the counterexample is to "hide" in such a zero a
second $\delta$-potential, which does not influence the previous ground state
but produces a new eigenvalue, which can be chosen to imply the above inequality. 

For the proof of the above theorem we consider at first the
operator $$H_l(c\delta_0) = (-\partial^2)^l - c\delta_0$$ with $l \in \mathbb{N}$
and $c \in (0,\infty)$, generated by the closure of the quadratic form
$$h_l(c\delta_0)[u,u] := \int_{\mathbb{R}} |\partial^l u|^2\,\mathrm{d}x -
c|u(0)|^2\quad\mbox{for}\;u \in \mathrm{C}_0^{\infty}(\mathbb{R}).$$
Let us mention without proof, that the domain of $H_l(c\delta_0)$ consists of all functions 
$u \in W^{2,2l-1}(\mathbb{R})\cap W^{2,2l}(\mathbb{R} \backslash \{0\})$ for which
\begin{equation*}
  \partial^{2l-1}u(0+) - \partial^{2l-1}u(0-) = (-1)^l\,c\,u(0).
\end{equation*}

As the computation in Appendix A shows, the operator $H_l(c\delta_0)$ has exactly one negative
eigenvalue $-\varkappa$ which satisfies
\begin{equation}\label{eq:kappal0}
  \varkappa^{\nu} =  L^0_{l,\nu,1}\,c.
\end{equation}

\begin{lemma}
  For $l \in \mathbb{N}$ with $l \geq 2$ the eigenfunction $u$ corresponding to the eigenvalue
  $-\varkappa$ of $H_l(c\delta_0)$ has at least one zero $x_0 \neq 0$. 
\end{lemma}

\begin{proof}[\textbf{Proof.}]
  Let us assume that $u$ has no zero. Then $(-\partial^2)^l u$ has, on the strength of the eigenvalue 
  equation
  \begin{equation*}
    (-\partial^2)^lu(x) = -\varkappa u(x)\quad\mbox{for}\quad x \in (-\infty,0),
  \end{equation*}
  no zero in $(-\infty,0)$ either.
  Because of
  \begin{equation*}
    \partial^{2l-1}u(x) = \int_{-\infty}^x \partial^{2l}u(t)\,\mathrm{d}t\quad\mbox{for}\quad x \in (-\infty,0)
  \end{equation*}
  the same holds for the function $\partial^{2l-1}u$, and so on for
  all lower derivatives up to the second derivative $u''$.
  It therefore follows from continuity of $u''$ that 
  \begin{equation}\label{eq:primezero}
    u'(0) = \int_{-\infty}^{0}u''(x)\,dx \neq 0.
  \end{equation}
  On the other hand $u$ is symmetric, which follows from the symmetry of the
  eigenvalue problem and the uniqueness (modulo a factor) of the
  eigenfunction. This together with the continuity of $u'$ implies $u'(0) =
  0$ in contradiction to \eqref{eq:primezero}. So $u$ has a zero in  $(-\infty,0)$.
\end{proof}
\noindent
One can indeed prove that $u$ has countably many zeros, by computing
$u$ explicitly. But with the existence of one zero we are already able to
prove Theorem \ref{theo:lbiggerl0}:

\begin{proof}[\textbf{Proof.}]
  Let $x_0 \neq 0$ be a zero of the eigenfunction $u_1$ corresponding
  to the unique negative eigenvalue $-\varkappa_1$ of 
  $H_l(\delta_0) = (-\partial^2)^l - \delta_0$. Because of
  \eqref{eq:kappal0} we have
  \begin{equation}
    \varkappa_1^{\nu} = L^0_{l,\nu,1}.
  \end{equation}
  For $\alpha > 1$ we consider the operator 
  \begin{equation*}
    H_l^{\alpha} = (-\partial^2)^l - \delta_0 - \alpha\delta_{x_0},
  \end{equation*}
  given by an appropriate quadratic form, which has exactly two negative
  eigenvalues. The latter follows from a standard variational argument.
  Obviously, $u_1$ is an eigenfunction of $H_l^{\alpha}$ corresponding to the
  eigenvalue $-\varkappa_1$. For the ground state of $H_l^{\alpha}$, which we refer to as $-\varkappa_0$,
  the inequality 
  \begin{equation}\label{eq:kappa2}
    \varkappa_0^{\nu} \geq L^0_{l,\nu,1}\,\alpha
  \end{equation}
  must hold. This is a consequence of the variational principle: If
  $\phi$ with $\|\phi\| = 1$ is the eigenfunction corresponding to the 
  unique negative eigenvalue
  $-\tau$ of the operator $H_l(\alpha\delta_{x_0}) = (-\partial^2)^l - \alpha\delta_{x_0}$ and if
  $h_l(\alpha\delta_{x_0})$ and $h_l^{\alpha}$ are the quadratic forms associated
  with $H_l(\alpha\delta_{x_0})$ and $H_l^{\alpha}$, then we have
  \begin{equation}
    \begin{split}
      -\tau &= h_l(\alpha\delta_{x_0})[\phi,\phi] = \|\partial^l \phi\|^2 - \alpha|\phi(x_0)|^2 \\
      &\geq \|\partial^l \phi\|^2 - |\phi(0)|^2 - \alpha|\phi(x_0)|^2 = h_l^{\alpha}[\phi,\phi].
    \end{split}
  \end{equation}
  By the variational principle the lowest eigenvalue of $H_l^{\alpha}$ is
  lower or equal $-\tau$.
  Because of \eqref{eq:kappal0} we have $\tau^{\nu} =
  L^0_{l,\nu,1}\,\alpha$. Therefore \eqref{eq:kappa2} holds. Notice
  that $-\varkappa_1$ cannot be the ground state of the operator 
  $H_l^{\alpha}$ if $\alpha > 1$.
 
  Furthermore we get 
  \begin{equation}
    \varkappa_0^{\nu} >  L^0_{l,\nu,1}\,\alpha
  \end{equation}
  if we choose $\alpha > 1$ in such a way, that
  $\phi(0) \neq 0$ holds. This is possible since 
  $\phi(x) = c u_1\left(\alpha^{\frac{1}{2l-1}}(x-x_0)\right)$, for $x
  \in \mathbb{R}$ and some $c \in \mathbb{C}$.
  So for a proper $\alpha > 1$ we have $\phi(0) = 
  c u_1\left(-\alpha^{\frac{1}{2l-1}} x_0\right) \neq 0$, since
  $u_1$ has only a countable set of zeros.

  Thus, $H_l^{\alpha}$ has two negative eigenvalues $-\varkappa_0$ 
  and $-\varkappa_1$, which fulfil the inequality
  \begin{equation}
    \varkappa_0^{\nu} + \varkappa_1^{\nu} > L^0_{l,\nu,1}(1+\alpha).
  \end{equation}
  Because of the Lieb-Thirring inequality, extended by a standard
  argument to $\delta$-potentials, we have
  \begin{equation}
    \varkappa_0^{\nu} + \varkappa_1^{\nu} \leq L_{l,\nu,1}(1+\alpha)
  \end{equation}
  and therefore $L_{l,\nu,1} > L^0_{l,\nu,1}$.
\end{proof}

\section{Lieb-Thirring inequalities for Riesz means of eigenvalues for
  polyharmonic operators in higher dimensions}
\label{ltinequalitydimhigh}
In this section we apply the ideas of Laptev and Weidl from
\cite{L-W} to obtain results valid in dimensions greater than one.

Consider the following Weyl type asymptotics:
\begin{alignat*}{4}
&\lim_{\alpha \rightarrow +\infty} \frac{1}{\alpha^{\gamma + \frac{d}{2l}}}
\, \tr \left(\sideset{}{_{j=1}^d} \sum  
\left(-\partial_j^2\right)^{l} + \alpha V\right)_{-}^{\gamma} =\\
&= \iint_{\mathbb{R}^d\times\mathbb{R}^d} 
\left(\sideset{}{_{j=1}^d} \sum \xi_{j}^{2l} + V\right)_{-}^{\gamma} 
\frac{dx \, d\xi}{(2\pi)^{d}} = 
C_{l,\gamma,d} \int_{\mathbb{R}^d} V_{-}^{\gamma + \frac{d}{2l}} \, dx. 
\end{alignat*}
We shall need the following lemma concerning the constants $C_{l,\gamma,d}$;
the proof is basically a lengthy computation which shall not be presented here.
\begin{lemma}
The constants $C_{l,\gamma,d}$ appearing in the above semi-classical
limit obey the following identity
\begin{equation}
  \label{Cidentity}
C_{l,\gamma,d} = C_{l,\gamma + \frac{1}{2l}, d-1} \cdot C_{l,\gamma,1}.
\end{equation}
Furthermore, the constants $C_{l,\gamma,d}$ are explicitly given by 
\begin{equation}
C_{l,\gamma,d} = \frac{1}{2(2\pi l)^{d}} 
\,B\left(\gamma+1,\frac{d}{2l}\right) \left(\frac{2\pi^{\frac{d-1}{2}} 
\Gamma \left(\frac{1}{2l}\right)}{\Gamma \left(\frac{ld + 1 -
    l}{2l}\right)} \right)^{d},
\end{equation}
where $B$ is the Beta-function.  
\end{lemma}
For the operator $(- \Delta)^{l}$ the ideas of Laptev and Weidl cannot be used
directly, because there is no simple way to separate the variables in this
case. Therefore we consider first the operator $\sum_{j=1}^d
\left(-\partial_j^2\right)^l$. 
\begin{theorem}\label{theo:varsepresult}
  Let $V : \mathbb{R}^d \rightarrow \mathcal{B}\left(\mathcal{G}\right)$ be 
  an operator-valued function satisfying $V(x) = (V(x))^{*}$ and 
  $V(x) \in S_{1}\left(\mathcal{G}\right)$ for a.e. $x \in \mathbb{R}^{d}$ and
  such that $\tr (V_{-}(\cdot))^{\gamma + \frac{d}{2l}} \in 
  L^{1}\left(\mathbb{R}^{d}\right)$, for some $\gamma \geq 1-\frac{1}{2l}$, $l
  > 1$.
  Then the following inequality holds true:
  \begin{equation}\label{eq:varsepestimate}
    \tr \left(\sum_{j=1}^d \left(-\partial_j^2\right)^l \otimes
      \mathbf{1}_{\mathcal{G}} + V\right)_{-}^{\gamma} \leq 
    \left(\frac{c_{l}}{L^{\text{\upshape cl}}_{l,1-\frac{1}{2l},1}}
    \right)^{d} C_{l,\gamma,d} \int_{\mathbb{R}^{d}} 
    \tr (V_{-}(x))^{\gamma + \frac{d}{2l}} \, dx.
  \end{equation}
  In the case $l=2$ the constant on the right hand side can
  be replaced by $\left(\frac{4}{3^{1/4} \sqrt{2}}\right)^{d} C_{2,\gamma,d}$.
\end{theorem}
\begin{proof}
  Using Corollary \ref{Biharmoniccorollary} and Corollary
  \ref{Polyharmoniccorollary} the result follows directly by applying the
  technique from \cite{L-W}, section 3. 
\end{proof}

Because 
\begin{equation}
  \sum_{j=1}^d \left(-\partial_j^2\right)^l \otimes
  \mathbf{1}_{\mathcal{G}} \leq (- \Delta)^{l} \otimes \mathbf{1}_{\mathcal{G}}
\end{equation}
in quadratic form sense, estimate \eqref{eq:varsepestimate} is also valid for the polyharmonic
operator case with $\sum_{j=1}^d\left(-\partial_j^2\right)^l$ replaced
by $(- \Delta)^{l}$. This follows from the minimax principle. Consequently we achieve
\begin{theorem} \label{most-general}
  Let $V$ and $\gamma$ be as in Theorem \ref{theo:varsepresult}. 
  Then the following inequality holds true:
  \begin{equation}
    \label{general}
    \tr ((- \Delta)^{l} \otimes \mathbf{1}_{\mathcal{G}} + V)_{-}^{\gamma} \leq
    \left(\frac{c_{l}}{L^{\text{\upshape cl}}_{l,1-\frac{1}{2l},1}}
    \right)^{d} C_{l,\gamma,d} \int_{\mathbb{R}^{d}} 
    \tr (V_{-}(x))^{\gamma + \frac{d}{2l}} \, dx.
  \end{equation}
  Again, for the biharmonic operator the constant on the right hand side can
  be replaced by $\left(\frac{4}{3^{1/4} \sqrt{2}}\right)^{d} C_{2,\gamma,d}$.
\end{theorem} 

\textbf{Remark.} It is interesting to note that the proofs above
may be modified as to include the case when the operator
$\mathbf{1}_{\mathcal{G}}$ is replaced by some other operator 
$A$ acting in $\mathcal{G}$. More precisely, let A be any
self-adjoint operator acting in $\mathcal{G}$ which is positive,
i.e.
\begin{equation*}
\langle A x, x\rangle_{\mathcal{G}}\,>\,0, \quad \mbox{for all } 0 \neq x \in \mathcal{G}.
\end{equation*}  
Then the inequality 
\begin{equation}
  \label{evenmoregeneral...}
\tr ((- \Delta)^{l} \otimes A + V)_{-}^{\gamma} \leq
\left(\frac{c_{l}}{L^{\mbox{\footnotesize\upshape cl}}_{l,1-1/2l,1}}
\right)^{d} C_{l,\gamma,d} \int_{\mathbb{R}^{d}} 
\tr  A^{-\frac{1}{2l}} (V_{-}(x))^{\gamma + \frac{d}{2l}} \, dx
\end{equation}
is valid whenever the right-hand side is finite, for $\gamma > 1 -
1/2l$, $l > 1$. Note that if $A$ is positive definite, i.e. 
\begin{equation*}
0 < m_{A} := \inf_{||x||=1}\,\langle Ax, x\rangle_{\mathcal{G}},
\end{equation*} 
then the bound \eqref{evenmoregeneral...} is valid for any $V$ satisfying 
the criteria listed in Theorem \ref{most-general}.
The proof of \eqref{evenmoregeneral...} is basically the same as that
of \eqref{general}. In the estimate from above, the scalar inequality 
\eqref{scalarinequality} is replaced by the operator-inequality 
\begin{equation*}
  \label{operator-inequality}
\frac{\epsilon^{2l - 1}}{B^{2l} + \epsilon^{2l}} \leq  
\tilde{c}_{l}\,\frac{\epsilon}{B^2 + \epsilon^2},
\end{equation*}
valid for any positive self-adjoint operator $B$ acting on $\mathcal{G}$. 
The majorization as well as the Aizenman-Lieb 
argument works out similar as before, as does the 
``lifting'' to dimensions greater than one.\\
Note that the same technique, applied to the special case of the operator 
\begin{equation}
  -\frac{d^2}{dx^{2}} \otimes A + V
\end{equation}
acting in $L^{2}\left(\mathbb{R}, \mathcal{G}\right)$, implies that
\begin{equation}
  \label{Schr-bound}
\tr \left(-\frac{d^2}{dx^{2}} \otimes A + V\right)_{-}^{\gamma} \leq
2\,L^{\mbox{\footnotesize\upshape cl}}_{1,\gamma, 1}
\int_{\mathbb{R}} 
\tr  A^{-\frac{1}{2}} (V_{-}(x))^{\gamma + \frac{1}{2}} \, dx,
\end{equation}
for any $\gamma \geq 1/2$. It is tantalizing to ask for the smallest
bound in \eqref{Schr-bound}. Does it, as in case $A =
\mathbf{1}_\mathcal{G}$ (see \cite{L-W}), hold with the classical constant
if we consider Riesz means of order $\gamma \geq 3/2$? This problem is
still open.

\appendix

\section{}

\begin{lemma}
  The unique negative eigenvalue $-\varkappa$ of the operator
  $H_l(c\delta_0) = (-\partial^2)^l - c\delta_0$ satisfies
  the identity
  \begin{equation}
    \varkappa^\nu = L_{l,\nu,1}^0 c.
  \end{equation}
\end{lemma}
\begin{proof}[\textbf{Proof.}]
  At first we notice, that $(-\partial^2)^l u = -\varkappa u$
  has the basic solutions
  \begin{equation*}
    g_k(x) := \exp(r_k\sqrt[2l]{\varkappa}\, x)\quad\mbox{for}\quad k=0,\dots,2l-1.
  \end{equation*}
  Here the $r_k$'s are the $2l$ complex roots of the equation $r_k^{2l} =
  (-1)^{l+1}$. It holds
  \begin{equation*}
    r_k = \exp\left(\frac{2k+1-l}{2l}i\pi\right)\quad\mbox{for}\quad k=0,\dots,2l-1.
  \end{equation*}
  Notice that the roots are ordered so that $r_0$ to
  $r_{l-1}$ have positive and $r_l$ to $r_{2l-1}$ have negative real
  parts. Therefore the functions $g_0$ to $g_{l-1}$ are
  not square integrable on $(0,\infty)$, and neither are $g_l$ to
  $g_{2l-1}$ on $(-\infty,0)$. Let us write $\mathfrak{g}(x) =
  \big(g_0(x),\dots,g_{2l-1}(x)\big)$ and $\partial^k\mathfrak{g}(x) =
  \big(\partial^kg_0(x),\dots,\partial^kg_{2l-1}(x)\big)$. Further let
  \begin{equation*}
    \mathfrak{G}(x) := \begin{pmatrix} \mathfrak{g}(x)\\ \partial\mathfrak{g}(x)\\ \vdots \\ \partial^{2l-1}\mathfrak{g}(x)\\ \end{pmatrix} \quad\mbox{and}\quad
    \mathfrak{E}(x) := \begin{pmatrix} 0\\ \vdots \\ 0 \\\mathfrak{g}(x) \end{pmatrix}.
  \end{equation*}
  Then we can formulate the conditions on the eigenfunction, which on
  each of the intervals $(-\infty, 0),(0,\infty)$ is a linear
  combination of the basic solutions, at the point $0$ as follows:
  \begin{equation*}
    \mathfrak{G}(0)\mathfrak{h}-\mathfrak{G}(0)\mathfrak{v} =  (-1)^l c \mathfrak{E}(0)\mathfrak{v}.
  \end{equation*}
  Here $\mathfrak{v},\mathfrak{h} \in \mathbb{C}^{2l}$ are the
  coefficients of the basic solutions on the left and
  right interval. Note, that the matrix $\mathfrak{G}(0)$ is invertible,
  because its determinant, the so called Wronskian determinant,
  is non-zero. Therefore the latter equation takes the form
  \begin{equation}\label{eq:matrixev}
    \big(I + (-1)^l c \mathfrak{G}^{-1}(0)\,\mathfrak{E}(0)\big) \mathfrak{v} = \mathfrak{h},
  \end{equation}
  where $I$ is the identity matrix.
  The inverse of the matrix
  \begin{equation*}
    \mathfrak{G}(0) = \big[g_k^{(n)}(x)\big]_{\begin{subarray}{l}n =
    0,\dots,2l-1\\k=0,\dots,2l-1 \end{subarray}} 
    = \left[(r_0\sqrt[2l]{\varkappa})^n\exp\left(\frac{nk}{l}i\pi\right)\right]_{\begin{subarray}{l}n = 0,\dots,2l-1\\k=0,\dots,2l-1 \end{subarray}}
  \end{equation*}
  is given by
  \begin{equation*}
    \mathfrak{G}^{-1}(0) =
        \frac{1}{2l}\left[\frac{\exp\left(\frac{(2l-n)k}{l}i\pi\right)}{(r_0\sqrt[2l]{\varkappa})^k}\right]_{
          \begin{subarray}{l}n = 0,\dots,2l-1\\ k=0,\dots,2l-1 \end{subarray}}.
  \end{equation*}
  We furthermore get, with $\tau := (-1)^l
  c(2l)^{-1}(r_0\sqrt[2l]{\varkappa})^{1-2l}$, that
  \begin{equation*}
    (-1)^lc\mathfrak{G}^{-1}(0)\,\mathfrak{E}(0) =
    \tau\left[\exp\left(\frac{n}{l}i\pi\right)\right]_{\begin{subarray}{l}n
    = 0,\dots,2l-1\\ k=0,\dots,2l-1 \end{subarray}} = \tau \begin{bmatrix} ye^T &
    ye^T\\ -ye^T & -ye^T \end{bmatrix},
  \end{equation*}
  where $y,e \in \mathbb{C}^l$ with $y := (\exp(\frac{0}{l}i\pi),\dots, \exp(\frac{l-1}{l}i\pi))^T$ and
  $e := (1,\dots,1)^T$.
  Notice now, that we have $\mathfrak{v}_l = \dots = \mathfrak{v}_{2l-1} = 0$
  and $\mathfrak{h}_0 = \dots = \mathfrak{h}_{l-1} = 0$, since the eigenfunction must be
  square integrable. Therefore, writing $\mathfrak{v} = \binom{\tilde{\mathfrak{v}}}{0}, \,\mathfrak{h} =
    \binom{0}{\tilde{\mathfrak{h}}}$, where
  $\tilde{\mathfrak{v}},\tilde{\mathfrak{h}} \in \mathbb{C}^l$
   we see that equation \eqref{eq:matrixev} has a non-trivial solution if and only if
  \begin{equation*}
    (I + \tau ye^T)\tilde{\mathfrak{v}} = 0
  \end{equation*}
  has a non-trivial solution. But it is not difficult to see, that the
  latter holds if and only if 
  \begin{equation*}
    \tau e^Ty+1 = 0,
  \end{equation*}
  that is if
  \begin{equation*}
    \varkappa^\nu =
    \left(\frac{(-1)^{(l+1)}}{2lr_0^{(2l-1)}}\sum_{k=0}^{l-1}
    \exp\left(\frac{k}{l}i\pi\right)\right)c = \frac{1}{2l\sin(\pi/(2l))}c =
    L_{l,\nu,1}^0 c.
  \end{equation*}
  This completes the proof.
\end{proof}

\section*{Acknowledgements}

The authors would like to thank Ari Laptev and Timo Weidl, who once introduced them to
this interesting topic. Their guidance is gratefully acknowledged. The first
author also thanks for the support by ESF SPECT and DAAD-SI, PPP-programme,
while the second author is deeply grateful to The Wenner-Gren Foundations for
their financial support.

{\noindent \small \textsl{e-mail:}\\
  \mbox{\texttt{foerster@mathematik.uni-stuttgart.de, ostensson@cims.nyu.edu}}
\end{document}